\documentclass[12pt,epsfig]{article}
\usepackage{graphicx}
\usepackage{amssymb}
\newcommand{\beq}[1] {\begin{equation}\label{#1} }
\newcommand{\eeq} {\end{equation} }

\newcommand{\bea}[1]{\begin{eqnarray}\label{#1} }
\newcommand{\eea}{\end{eqnarray}}

\newcommand{\hmu}{{\hat\mu}}
\newcommand{\hnu}{{\hat\nu}}

\newcommand{\hh}{{\hat{h}}}

\newcommand{\tA}{{\widetilde{A}}}

\newcommand{\tih}{{\widetilde{h}}}
\newcommand{\tp}{{\widetilde\phi}}

\newcommand{\vn}{{\vec{n}}}

\newcommand{\gsim}{\lower.7ex\hbox{$\;\stackrel{\textstyle>}{\sim}\;$}}
\newcommand{\lsim}{\lower.7ex\hbox{$\;\stackrel{\textstyle<}{\sim}\;$}}
\newcommand{\del}{\partial}

\newcommand\lrpd{\partial^{\!\!\!\!\! ^\leftarrow\!\!\!\! ^\rightarrow}}

\newcommand{\eps}{\epsilon}
\newcommand{\al}{\alpha}
\newcommand{\be}{\beta}
\newcommand{\ga}{\gamma}
\newcommand{\si}{\sigma}
\newcommand{\ro}{\rho}
\newcommand{\la}{\lambda}
\newcommand{\de}{\delta}
\begin{document}

\vspace*{-0.5cm}
\begin{flushright}
OSU-HEP-03-7
\end{flushright}
\vspace{0.5cm}

\begin{center}
{\Large
{\bf Gravity and Matter in Extra Dimensions} }

\vspace*{1.5cm}
 C. Macesanu\footnote{Email address:  mcos@pas.rochester.edu}$^{,\dag}$,
A. Mitov\footnote{Email address: amitov@pas.rochester.edu}$^{,\ddag}$
and S. Nandi\footnote{Email address: shaown@okstate.edu}$^{,\dag}$

\vspace*{0.5cm}
$^{\dag}${\it Department of Physics, Oklahoma State University\\
Stillwater, Oklahoma, 74078\\}

$^{\ddag}${\it Department of Physics and Astronomy, University of Rochester\\
Rochester, NY, 14627.}
\end{center}

\begin{abstract}

In this paper, we derive from the viewpoint of the
effective 4D theory the interaction terms between linearized gravity
propagating in $N\geq 2$ large extra dimensions and matter propagating
into one extra dimension.
This generalizes known results for the interactions 
between gravity and 4D matter in ADD-type models.
Although we assume that matter is described by an
Universal Extra Dimensions (UED) scenario (with all fields
propagating into the fifth dimension), we present our
results in a general form that can be easily adapted to
various other scenarios of matter distribution.
We then apply our results to the UED 
model on a fat brane and consider some
phenomenological applications. Among these are the computation of the
gravitational decay widths of the matter KK excitations and the effect the
width of the brane has on the interactions of gravity with Standard Model
particles. We also estimate the cross-section for producing single KK
excitations at colliders through KK number-violating gravitational
interaction.

\end{abstract}

\section{Introduction}

At present, various theoretical constructions (most notably string theory) 
imply that there may exist
more space-time dimensions than the four directly accessible
to our senses. For a long time, it was assumed that such extra
dimensions are compactified on a scale of order $1/M_{Pl}$, and
thus inaccessible to the present day colliders, or those 
of the foreseeable future. In
light of recent ideas \cite{witten}, however, it is conceivable
that the compactification radius of these extra dimensions is
close to the inverse TeV scale or larger, which may have
implications for the phenomenology of current day colliders.

There are several types of models with extra-dimensions that can
lead to interesting low-energy  physics. Relevant for our paper
are the Arkani-Hamed--Dimopoulos--Dvali (ADD) type of models
\cite{ADD}, which exhibit factorizable geometry, and where the extra
dimensions form a compact manifold. Each field that can propagate
beyond the four non-compact dimensions (the bulk) can then be
expanded in Kaluza-Klein (KK) modes, and manifests itself in the
4D effective theory as a tower of fields with increasing mass, the
interval between the masses of two consecutive excitations being
of order $1/r$.

The ADD scenario was initially conceived as a way to explain the
hierarchy between the electro-weak scale and the Planck scale. The
reason for gravity being so weak is that its strength is diluted
by propagating in extra dimensions. The ADD relation between the
4D Planck scale $M_{Pl}$ and the fundamental Planck scale $M_D$ is
given by
$$ M_{Pl}^2 = M_D^{N-2} r^N,$$
which implies that $M_D$ can be as low as TeV,
for values of the compactification scale $r$ of order
eV$^{-1}$ up to keV$^{-1}$, depending on the number of extra dimensions $N$.

In the original ADD model, only gravity propagates in the extra
dimensions, while matter is confined to a 4D brane. One has also
the option of having the ordinary matter fields propagate in the
bulk. There are a variety of models of this type. In some of them
just a certain set of fields (usually the gauge bosons or the
Higgs fields) propagate in extra dimensions \cite{gb}; in other,
called Universal Extra Dimensions (UED) models, all Standard Model
fields propagate into the bulk \cite{ACD}. They also differ in the
number of dimensions in which matter propagates; usually this is
taken to be 5 or 6, but can conceivably be higher.

Note, however, that having matter in the extra dimensions
introduces a certain asymmetry in the model. Experimental
non-observation of KK matter excitations requires that the radius
of the space on which matter is compactified be of order of at
least inverse TeV. This is much smaller than the eV$^{-1}$ (or
keV$^{-1}$) size needed to achieve the natural scale for gravity
as in the ADD scenario. One can think of several ways in which to
accommodate such bound: first, one can assume that in such a model
the space is asymmetric; the one or two extra dimensions in which
matter propagate have radius of inverse TeV size, while the
dimensions in which only gravity propagates are of inverse eV size
\cite{asym}. Another possibility is to take the size of the whole
space to be of the order inverse eV, but to assume that the matter
fields are confined by some mechanism close to the 4D brane. This
is known as the fat brane scenario \cite{rigolin}, where the
length $R$ in which matter can propagate into the bulk is the
width of the brane.

Besides answering some long standing theoretical questions, models
where gravity and possibly matter propagate in extra dimensions
may also have very interesting phenomenological consequences.
Although the probability of radiating a specific KK graviton is
very small, there is a large number of them, which can lead to
measurable effects in collider processes \cite{hewett, rizzo,
direct_prod}. In models with matter propagating in extra
dimensions, gravity can play additional roles. For example, in
UED-type models, it can mediate the decays of stable first level
KK excitations, thus leading to interesting collider signals
\cite{cmn,cmn2}.

The purpose of this paper is to derive the interactions of gravity
with matter in the 4D effective theory for the case when both
propagate in extra dimensions. These interactions have been
derived for the case when matter is restricted to a 4D brane
\cite{HLZ, GRW}. We generalize here these results for models where
matter propagates in one compact extra dimension. Although our
results are applicable for a larger class of models, for the sake
of specificity, we formulate them in an UED scenario with matter
living on a fat brane. We then analyze some phenomenological
implications of the gravity-matter interactions in this particular
model.

The paper is organized as follows: in the next section we describe
concisely the model under consideration, i.e. we separately review
the UED scenario for matter propagating in one extra dimension, as
well as the KK reduction of linearized gravity on an
$N$-dimensional torus. This allows us to introduce our conventions
and notations. In Section 3 we derive the form of the interaction
terms between matter and gravity in the 4D effective theory and
also present the corresponding Feynman rules. (The Feynman rules for the
interactions involving more than two matter fields are presented
in the Appendix).
Section 4 deals with
phenomenological applications, including the computation of
gravitational decay widths of KK excitations of matter, and the
possibility of producing single KK excitations at colliders
through gravity mediated interactions. At the end we present our conclusions.

\section{Gravity and matter in extra dimensions}

We start by specifying the parameters of our model. Gravity is 
assumed to propagate in $N$ `large' extra dimensions compactified
on a torus,
which have common size of inverse eV up to inverse keV. We denote
the radius of these dimensions by $r/2 \pi$; then the linearized metric
field has a KK expansion:
\beq{grav_exp} \hh_{\hmu\hnu}(x,y)\ =\ \sum_{\vec n}
\hh_{\hmu\hnu}^{\vec{n}}(x)\ \exp\left(i {2\pi
\vec{n}\cdot\vec{y}\over r}\right)\ . \eeq
With respect to gravity, mostly the notations introduced in
\cite{HLZ} are used; that is, the `hat' denotes quantities which
live in 4+$N$ dimensions: $\hmu, \hnu = 0,\ldots,3,5,\ldots 4+N$,
while $\mu, \nu = 0,\ldots,3$. The $(4+N)$D graviton field is
decomposed into 4D tensor, vector and scalar components by:
\beq{hh_def} \hh_{\hmu\hnu}\ =\ V_N^{-1/2}\left(\begin{array}{cc}
h_{\mu\nu}+\eta_{\mu\nu}\phi & A_{\mu i}\\
A_{\nu j}   &  2 \phi_{ij}
\end{array}\right)\
\eeq
where $V_N= r^N$ is the volume of the $N$-dimensional torus. 
Also, $i,j = 5,6,\ldots 4+N$ and $\phi = \phi_{ii}$. The
fields $h_{\mu\nu}, A_{\mu i}$ and $\phi_{ij}$ are decomposed in
KK modes as in Eq. (\ref{grav_exp}).

We take the matter to propagate into one extra dimension,
considered to be the fifth one. Since keV spaced excitations of
the SM fields have not been observed, we  have to restrict the
matter on a fat brane \cite{rigolin}; that is, the matter fields
can go only a limited length $\pi R$ in the fifth dimension, with
$R$ of order inverse TeV. Moreover, in order to obtain chiral 4D
fermions, it is necessary to impose additional constraints on the
extra-dimensional fields. Usually this is achieved by placing the
higher dimensional fermions on an orbifold $S^1/\mathbb{Z}_2$ of
length $\pi R$. This is also the case for the fermions in the UED
model. The matter fields are expanded in KK modes as follows \cite{cmn}: 
 \bea{sm_exp} (\Phi, B_{\mu}^a)  & = &
\frac{1}{\sqrt{\pi R}}\left[ (\Phi_0, B_{\mu, 0}^a)  +
\sqrt{2} \sum_{n=1}^{\infty} (\Phi_n, B_{\mu ,n}^{a})  \cos(\frac{n y}{R}) \right]  \nonumber \\
Q  &  = &
\frac{1}{\sqrt{\pi R}} \left\{ Q_{L} + \sqrt{2} \sum_{n=1}^{\infty}
\left[ Q_L^n  \cos \left(\frac{n y}{R} \right)
 + Q_R^n  \sin \left(\frac{n y}{R} \right) \right] \right\} \nonumber \\
q  &  = &
\frac{1}{\sqrt{\pi R}} \left\{ q_{R} + \sqrt{2} \sum_{n=1}^{\infty}
\left[ q_R^n  \cos \left(\frac{n y}{R} \right)
 + q_L^n  \sin \left(\frac{n y}{R} \right) \right] \right\}.
\eea
Here, $\Phi$ and $B_{\mu}^a$ are scalar and vector boson fields;
 the gauge for the latter ones can be chosen such that $B_5^a = 0$
\cite{gauge_fix}. $Q$ and $q$ are the two fermion fields in 5D corresponding
to each fermion field in the SM, which are doublets
($Q$) respectively singlets
($q$) under the $SU(2)$ gauge group, and whose zero modes are the left,
respectively right-handed components of the SM fermion fields.

The KK decomposition of the fields in Eq.(\ref{sm_exp}) ensures
that the 5D fields 
have definite parity with respect to
reflections $y\to -y$ of the co-ordinate $y:\ -\pi R\leq
y\leq \pi R$ that parameterizes the circle $S^1$. Alternately, as it
was noted in \cite{ACD}, the decompositions (\ref{sm_exp})
correspond to the KK expansion of a 5D field restricted to the
interval $0\leq y \leq \pi R$ with the field and its derivatives
satisfying certain boundary conditions on the end-points $y=0,\pi
R$. The above discussion suggests the following possibilities for
the embedding of the fat brane into the fifth dimension: The first
one is to take the interval $0\leq y \leq \pi R$ to be a (small)
segment of the circle that parameterizes the fifth dimension. The
matter fields obey boundary conditions at the end-points of the interval,
and are then decomposed as in Eq.(\ref{sm_exp}). Gravity
is unaffected by the orbifolding of the matter, i.e. it contains
both even and odd modes to which the matter couples. Such an
approach was used in \cite{rizzo, cmn} and we shall adopt it in
this paper as well. The other possibility is that the whole fifth
dimension is an orbifold of length $r/2$ and the matter is
embedded as an interval of length $\pi R$. The other extra
dimensions that are generally assumed to form a torus may or may
not be subject to such additional discrete symmetry. The resulting
model describes the same matter content as the first one but now
the gravity components will also have specific parity i.e. the
decomposition in (\ref{grav_exp}) as well as the gravity-matter
interactions will be altered.

Since in this paper we do not attempt to address the question of
the origin of such fat brane construction, it is appropriate to
make the following short comment: although it may be less
realistic, the first model is easier to deal with in practice
because gravity is not affected by orbifolding. Hence, we will use this 
scenario in the following computations.
 However, our results can
be applied for the second scenario as well; one needs in
addition to project out the gravity modes with unwanted parity and
to modify the form-factors introduced in Eq. (\ref{ff}) bellow.

The decomposition in Eqns. (\ref{sm_exp}) ensures that the kinetic
terms of the matter KK excitations in the effective theory have
their canonical form in terms of the $\Phi_n, \psi_n$ and $B_n$
fields. This does not hold for the gravity Lagrangian when written
in terms of the fields $h_{\mu\nu}^\vn, A_{\mu i}^\vn$ and
$\phi_{ij}^\vn$. It is then necessary to redefine the fields in
the gravitational sector and to work in terms of `physical' fields
$\tih_{\mu\nu}^\vn, \tA_{\mu i}^\vn$ and $\tp_{ij}^\vn$ that have
canonical kinetic and mass terms. The details of this redefinition
are worked out in \cite{HLZ,GRW}. Here we shall just review the
results.

The $D (D-3)/2$ internal degrees of freedom (d.o.f) of a $D$ dimensional
massless graviton ($D = 4+N$) will appear, after dimensional
reduction, as the components of a massive spin 2 graviton
$\tih_{\mu\nu}^\vn$ (5 d.o.f), $N-1$ massive vector bosons
$\tA_{\mu i}^\vn$, with $ n_i \tA_{\mu i}^\vn =0 $ (3 d.o.f.
each), and $N (N-1)/2$ massive scalars $\tp_{ij}^\vn$, with $n_i
\tp_{ij}^\vn = 0$ and $\tp_{ij}^\vn = \tp_{ji}^\vn $ (1 d.o.f.
each). All of these fields have the same mass $m_\vn^2 = 4 \pi^2
\vn^2 / r^2$. The gauge of the spin-two and spin-one physical
fields is fixed by $\del^\mu \tih_{\mu\nu}^\vn = 0$,
$\tih_{\mu}^{\vn \mu} = 0$, and $\del^\mu \tA_{\mu i}^\vn =0 $.

The definition of the physical `tilde' fields in terms of the initial
fields $h_{\mu \nu}, A_{\mu i}$ and $\phi_{ij}$ is given in \cite{HLZ}.
We will need in the following to express the initial fields in terms
of the physical fields. In order for this to be possible, we have to fix
the gauge for the initial fields, too. There are $D (D+1)/2$ degrees
of freedom in $\hh_{\hmu \hnu}$; the de Donder gauge condition
\beq{donder}
\partial^\hmu (\hh_{\hmu\hnu}-{1\over2}\eta_{\hmu\hnu}\hh) =0
\eeq
eliminates $D$ d.o.f.; we can choose the additional conditions at
each KK level as:
\beq{gauge_fix}
n_i A_{\mu i}^\vn = 0,  \ \ n_i \phi_{ij}^\vn = 0
\eeq
which will eliminate the other spurious $D$ d.o.f. With these definitions:
\bea{h_def}
h_{\mu \nu}^\vn & = & \tih_{\mu \nu}^\vn +
\omega \left( \frac{ \del_\mu \del_\nu}{m_\vn^2} - {1\over 2}
\eta_{\mu \nu}  \right)  \tp^\vn \\
A_{\mu i}^\vn & = & \tA_{\mu i}^\vn \\
\phi_{ij}^\vn & = & {1 \over \sqrt{2}} \tp_{ij}^\vn -
{ 3 \omega a\over 2 }
 \left( \delta_{ij}- { n_i n_j \over \vn^2} \right) \tp^\vn
\eea
and $\phi^\vn = (3\omega /2) \tp^\vn$. Here $\phi^\vn = \phi^\vn_{ii}$ (
same for the tilde fields),
$\omega = \sqrt{2/3 (N+2)}$, and
$a$ is a solution of the equation $3(N-1)a^2 + 6 a = 1$.

Now we have all the ingredients necessary to compute the interactions
of gravity with matter. Following the notations in \cite{HLZ}, the $D$
dimensional action is:
\beq{act_gen}
{\cal S}_{int} = -{{\hat \kappa} \over 2}
\int d^D x \ \delta(x^6) \ldots \delta(x^N)\
 \hh^{\hmu\hnu} T_{\hmu\hnu}
\eeq where $T_{\hmu\hnu}$ is the energy-momentum tensor of the 5D
matter. Expanding the gravity field in KK modes we obtain:
\beq{act_4d} {\cal S}_{int} = -{\kappa \over 2} \int d^4 x\
\int_0^{\pi R} dy \ \sum_{\vn} \left[ \left( h^\vn_{ \mu \nu} +
\eta_{\mu \nu} \phi^\vn \right) T^{\mu \nu} -2 A^\vn_{\mu 5}
T^\mu_5 + 2 \phi^\vn_{55} T_{55} \right] e^ {2\pi i { n_5 y\over
r}}. \eeq In terms of the physical fields, the effective 4D
Lagrangian is: \bea{l_4d} {\cal L}_{int} & = & -{\kappa \over 2}
\sum_{\vn} \int_0^{\pi R} dy \ \left\{ \left[ \tih^\vn_{ \mu \nu}
+ \omega \left( \eta_{\mu \nu} + \frac{\del_\mu \del_\nu}{m_\vn^2}
\right) \tp^\vn \right]
T^{\mu \nu}   - \right.  \nonumber \\
& & \left. 2 \tA^\vn_{\mu 5} T^\mu_5 \ + \
 \left(  \sqrt{2} \tp^\vn_{55} - \xi \tp^\vn \right) T_{55}
\right\} e^ {2\pi i { n_5 y\over r}}
\eea
where $\xi = 3 \omega a  ( 1- n_5^2 / \vn^2)$. Defining the projections of
the matter energy momentum tensor on the $\vn$-th graviton state by
$$ T_{MN}^{n_5} (x) =
\int_0^{\pi R} dy\ T_{MN}(x,y)\ e^ {2\pi i { n_5 y\over r}} $$
we can write
\beq{l_4d2}
{\cal L}_{int}  =   -{\kappa \over 2} \sum_{\vn}
\left\{ \left[ \tih^\vn_{ \mu \nu} +
\omega \left( \eta_{\mu \nu} + \frac{\del_\mu \del_\nu}{m_\vn^2} \right)
\tp^\vn \right]
T_{n_5}^{\mu \nu}   - \right.   \left. 2 \tA^\vn_{\mu 5} 
T_{n_5 5}^{~\mu}  +
 \left(  \sqrt{2} \tp^\vn_{55} - \xi \tp^\vn \right) T^{n_5}_{55}
\right\}. \eeq In the following section, we will give the 
explicit form of the
energy-momentum tensor for different types of matter, and 
the corresponding Feynman rules for the interaction of matter
with gravity.

\section{Gravity-matter interactions}

\subsection{Scalar matter}

The energy momentum tensor in 5D for a complex scalar  field is:
\beq{t_scal}
T^{ S}_{MN}\ =\ -\eta_{MN} ( D^R\Phi^\dagger D_R\Phi
        - m^2_\Phi\Phi^\dagger\Phi )
    +D_M\Phi^\dagger D_N\Phi
    +D_N\Phi^\dagger D_M\Phi\ ,
\eeq where $M,N,R$ are indices which run from 0 to 5, and $D_M =
\del_M + i g B^a_M T^a$ is the covariant derivative. We can expand
the fields in KK modes, and integrate over the fifth dimension, to
obtain the projections of the energy-momentum tensor: \bea{ts_4d}
T^{ S n }_{\mu \nu}  & = & \sum_{l,m} \biggl\{
 \left[ -\eta_{\mu \nu} ( \del^\rho\Phi^{\dagger m} \del_\rho\Phi^l
        - m^2_\Phi\Phi^{\dagger m} \Phi ^l)
    +\del_{(\mu}\Phi^{\dagger m} \del_{\nu)}\Phi^l
    \right] {\cal F}^{(cc)}_{lm|n}
    + \eta_{\mu \nu} {ml\over R^2} \Phi^{\dagger m} \Phi ^l
    {\cal F}^{(ss)}_{lm|n} \biggr\} \nonumber \\
 & & - i g \sum_{l,m,k}
 \left[ -\eta_{\mu \nu} \Phi^{\dagger m} \hat{B}^{k\rho} \lrpd_\rho\Phi^l
  + \Phi^{\dagger m} \hat{B}^k_{(\mu} \lrpd_{\nu)} \Phi^l
 \right] {\cal F}^{(ccc)}_{lmk|n}
         \nonumber \\
 & &         + g^2 \sum_{l,m,k,j} \biggl[ \Phi^{\dagger m}
                \biggl( -\eta_{\mu \nu} \hat{B}^{k\rho} \hat{B}^j_{\rho}
                 + \hat{B}^k_{(\mu} \hat{B}^j_{\nu)}
                \biggr) \Phi^l \biggr]
        {\cal F}^{(cccc)}_{lmkj|n}  \nonumber \\
T^{ S n  }_{\mu 5}  & = &  \sum_{l,m} \biggl\{
     (\del_\mu \Phi^{\dagger m} ) \Phi^l
        {l \over R} {\cal F}^{(sc)}_{lm|n}  +
        \Phi^{\dagger m} (\del_\mu \Phi^l) {m \over R} {\cal F}^{(cs)}_{lm|n}
        \biggr\}  \nonumber \\
        & & + i g  \sum_{l,m,k}  \Phi^{\dagger m} \hat{B}^k_\mu  \Phi ^l
        \biggl( -{l \over R} {\cal F}^{(scc)}_{lmk|n}
         + {m \over R} {\cal F}^{(csc)}_{lmk|n} \biggr) \nonumber \\
T^{ S n }_{5 5}  & = & \sum_{l,m} \biggl\{
( \del^\rho\Phi^{\dagger m} \del_\rho\Phi^l
        - m^2_\Phi\Phi^{\dagger m} \Phi ^l) {\cal F}^{(cc)}_{lm|n}
        + {ml\over R^2} \Phi^{\dagger m} \Phi ^l  {\cal F}^{(ss)}_{lm|n}
        \biggr\} \nonumber \\
        & & - i g \sum_{l,m,k}  \Phi^{\dagger m} \hat{B}^{k\rho}
  \lrpd_\rho\Phi^l {\cal F}^{(ccc)}_{lmk|n}
        + g^2 \sum_{l,m,k,j} \Phi^{\dagger m}
 \hat{B}^k_{\rho} \hat{B}^{j\rho} \Phi^l {\cal F}^{(cccc)}_{lmkj|n}
\eea In the above, $l,m,k,j$ are the KK excitation numbers of the
matter fields, with values starting from zero. We also use the
following notations: $X_{(\mu} Y_{\nu)} = X_\mu Y_\nu + X_\nu
Y_\mu$, $X \lrpd Y = X (\del Y) - (\del X) Y$ (in (\ref{ts_4d})
the derivatives act only on the scalar fields $\Phi$), and
$\hat{B} = B^a T^a$, with sum over the gauge index $a$. (We do not
write explicitly the color or SU(2) indices;
 terms like $\Phi^{\dagger m} \hat{B}^k \Phi ^l $ should
be read like $\Phi^{\dagger m}_\alpha T^a_{\alpha \beta} \Phi ^l_\beta
 B^{ak}$).
The form factors ${\cal F}^{(c...)}_{lm..|n}$ quantify the
superposition of the wave functions of the interacting fields (or
their derivatives) in the fifth dimension:
\beq{ff} {\cal F}^{(...
\hbox{f}_i...)}_{...l_i...|n}  =
 \int_0^{\pi R}
 dy \prod_i  c_i \hbox{f}_i \biggl( {l_i y\over R}\biggr) \
 \exp\biggl( 2 \pi i{n y \over r}\biggr) ,
\eeq
where the functions $\hbox{f}_i()$ can be sin() or cos(),
and $c_i = \sqrt{2/\pi R}$ if $l_i \ne 0$ or  $c_i = \sqrt{1/\pi R}$
for $l_i = 0$.

\subsection{Vector boson matter}

In 5D the energy momentum tensor for a vector boson field is:
\beq{t_vect} T^{ V }_{MN}  = \eta_{MN} \biggl({1\over4} F^{aRS}
F^a_{RS}-{m^2_B\over2} B^{aR} B^a_R \biggr) - \biggl( F_M^{a~R}
F^a_{NR} - m_B^2 B^a_M B^a_N \biggr) \eeq with $F^a_{RS} = \del_R
B^a_S -  \del_S B^a_R + g f^{abc} B^b_R B^c_S$ the field tensor in
five dimensions ($a$ is the gauge index of the field). For
purposes of generality, we added a 5D mass term in our
expressions.

Expanding (\ref{t_vect}) in KK modes and integrating over the
fifth dimension: \bea{tv_4d} T^{ V n }_{\mu\nu} &  = & \sum_{l,m}
\biggl\{  \left[ \eta_{\mu\nu}
    \biggl({1\over4} \tilde{F}^{am \rho\sigma}
    \tilde{F}^{al}_{\rho\sigma}-{m^2_B\over2} B^{am \rho} B^{al}_\rho\biggr)
    - \biggl( \tilde{F}_\mu^{am \rho} \tilde{F}^{al}_{\nu\rho}
    - m_B^2 B^{am}_\mu B^{al}_\nu \biggr) \right] {\cal F}^{(cc)}_{lm|n}
     \nonumber \\
& & - {ml\over R^2} \left( {1\over 2} \eta_{\mu\nu} B^{am \rho} B^{al}_\rho
       - B^{am}_\mu B^{al}_\nu \right) {\cal F}^{(ss)}_{lm|n}
        \biggr\}  \nonumber \\
& &  + g f^{abc} \sum_{l,m,k} \biggl\{
        B^{bm\rho} B^{cl\si} {\eta_{\mu\nu} \over 2}
        \tilde{F}^{ak}_{\rho\si} -   B^{bm}_{(\mu} B^{cl\rho}
           \tilde{F}_{\nu)\rho}^{ak}
    \biggr\}  {\cal F}^{(ccc)}_{lmk|n} \nonumber \\
& &     + g^2 f^{abc} f^{ade} \sum_{l,m,k,j} \biggl\{
    {\eta_{\mu\nu} \over 4} B^{bm\rho} B^{cl\si} B^{dk}_\rho B^{ej}_\si
    - B^{bm}_\mu B^{cl\rho} B^{dk}_\nu B^{ej}_\rho
    \biggr\}  {\cal F}^{(cccc)}_{lmkj|n} \nonumber \\
T^{ V n }_{\mu 5} &  = &  \sum_{l,m} \left(
     {l\over R}  \tilde{F}^{am \rho}_{\mu}
    B^{al}_\rho \right)  {\cal F}^{(sc)}_{lm|n}
    + g f^{abc} \sum_{l,m,k} \left( {l\over R} B^{bm}_\mu B^{ck}_\rho
    B^{al\rho} \right) {\cal F}^{(scc)}_{lmk|n}
     \nonumber \\
T^{ V n }_{55} &  = & -\sum_{l,m} \biggl\{ \biggl({1\over4}
    \tilde{F}^{am \rho\sigma}
    \tilde{F}^{al}_{\rho\sigma}-{m^2_B\over2} B^{am \rho} B^{al}_\rho\biggr)
    {\cal F}^{(cc)}_{lm|n} + {1\over 2} {ml\over R^2}
    B^{am \rho} B^{al}_\rho \ {\cal F}^{(ss)}_{lm|n} \biggr\}
    - {g\over 2} f^{abc} \nonumber \\
& &  \times \sum_{l,m,k}  \biggl(  B^{bm\rho} B^{cl\si}
    \tilde{F}_{\rho \si}^{ak} \biggr) {\cal F}^{(ccc)}_{lmk|n}
     - {g^2 \over 4} f^{abc} f^{ade} \sum_{l,m,k,j} \biggl(
     B^{bm\rho} B^{cl\si} B^{dk}_\rho B^{ej}_\si
     \biggr)  {\cal F}^{(cccc)}_{lmkj|n}
\eea Here $l,m,k,j$ are KK indices, and $\mu, \nu, \si, \rho$ are
4D space-time indices. Also, $\tilde{F}^{al}_{\rho\sigma} =
\del_\rho B^{al}_\si - \del_\si B^{al}_\rho$.

\subsection{Fermionic matter}

The energy momentum tensor for a fermion field in 5D is:
\bea{t_ferm} T^{ F }_{MN} & = & -\eta_{MN} \left(  \bar{\psi}i
\Gamma^R
 D_R \psi - {1\over 2} \del^R  (\bar{\psi} i \Gamma_R \psi)
 - m_\psi \bar{\psi} \psi \right)
\nonumber \\
 & & + \biggl( {1\over 2} \bar{\psi} i \Gamma_M D_N \psi
   -{1 \over 4} \del_M ( \bar{\psi} i \Gamma_N  \psi )
    + ( M \leftrightarrow N ) \biggr)
\eea with the covariant derivative $D_M = \del_M + i g B^a_M T^a$.
A 5D mass term is also added in here for purposes of generality.
The $\Gamma_M$ matrices in five dimensions are: $\Gamma_\mu =
\gamma_\mu, \Gamma_5 = i \gamma_5$. Expanding in KK modes for the
fields in (\ref{sm_exp}), we get: \bea{tfd_4d3} T^{ F n }_{\mu
\nu} & = & \sum_{l,m}  \bar{\psi}^m   \left[
    {i\over 4}\left[\gamma_{(\mu} \lrpd_{\nu)} -
    2\eta_{\mu\nu}\gamma^\sigma\lrpd_\sigma \right]
    {\cal G}^{(c)\mp}_{lm|n}
    + { \eta_{\mu \nu} \over 2 R}
    \left(  l  {\cal G}^{(c)\pm}_{lm|n}  +
     m {\cal G}^{(c)\mp}_{lm|n} \right)
     \right. \nonumber \\
 & & \left.  \pm   \eta_{\mu \nu} m_Q  {\cal G}^{(s)\pm}_{lm|n}
     \right] \psi^l
     + g \sum_{l,m,k}  \bar{\psi}^m   \left[
     \biggl( \eta_{\mu\nu} \gamma^\rho \hat{B}_\rho^k
     - {1 \over 2} \gamma_{(\mu} \hat{B}_{\nu)}^k \biggr)
    {\cal G}^{(cc)\mp}_{lmk|n} \right] \psi^l    \nonumber \\
T^{ F n }_{\mu 5} & = &  \sum_{l,m} \bar{\psi}^m  {1\over 4} \left[
    - \lrpd_\mu  {\cal G}^{(s)\pm}_{l,-m|n}
    + {i \gamma_\mu \over R} \left( l  {\cal G}^{(s)\mp}_{l,-m|n}  +
     m  {\cal G}^{(s)\pm}_{l,-m|n}
    \right) \right]\psi^l \nonumber \\
 & & - i  g \sum_{l,m,k}  \bar{\psi}^m   {1\over 2} \left[
     \hat{B}_\mu^k \ {\cal G}^{(sc)\pm}_{l,-mk|n}  \right] \psi^l
    \nonumber \\
T^{ F n }_{5 5} & = & \sum_{l,m}  \bar{\psi}^m   \left[
    {1\over 2} i \gamma^\rho \lrpd_\rho {\cal G}^{(c)\mp}_{lm|n}
    \mp m_Q  {\cal G}^{(s)\pm}_{lm|n}
    \right]\psi^l   - g \sum_{l,m,k}  \bar{\psi}^m   \left[
      \gamma^\rho \hat{B}_\rho^k \
    {\cal G}^{(cc)\mp}_{lmk|n} \right] \psi^l .
\eea
Here $\psi$ stands for either the doublet or singlet fields in
Eqs. (\ref{sm_exp}); the upper sign applies for the doublet case (that is,
for $\psi^m = Q^m_L + Q^m_R$) and the lower sign
applies for the singlet case ($\psi^m = - q^m_L + q^m_R$, the minus sign here
being due to a $\gamma_5$ rotation necessary to obtain the
canonical form for the mass terms). For the Standard Model in extra dimensions,
we have a doublet $(u_L, d_L)$ and two singlet set of fields
$u_R, d_R$ for each generation.
(Note that $\ga_5$ and $m_Q$ both change their signs when going from
the doublet to the singlet fields). In the limit where
the Yukawa interactions are negligible \cite{ACD, cmn}, the fields
$\psi^m$ are the mass eigenstates for the fermion KK excitations.

 The following form factors are used
in the above expressions:
\bea{g_def}
{\cal G}^{(c)\pm}_{lm|n} =
     ({\cal F}^{(c)}_{l-m|n} \pm {\cal F}^{(c)}_{m+l|n} \gamma_5 )/2
& , & {\cal G}^{(cc)\pm}_{lmk|n} =
     ({\cal F}^{(cc)}_{l-m,k|n} \pm {\cal F}^{(cc)}_{m+l,k|n} \gamma_5 )/2
     \nonumber \\
{\cal G}^{(s)\pm}_{lm|n} =
     ({\cal F}^{(s)}_{m+l|n} \pm {\cal F}^{(s)}_{l-m|n}  \gamma_5 )/2
& , & {\cal G}^{(sc)\pm}_{lmk|n} =
     ({\cal F}^{(sc)}_{m+l,k|n} \pm {\cal F}^{(sc)}_{l-m,k|n}  \gamma_5 )/2
\eea In a slight abuse of notation, the form factors ${\cal
F}^{(c),(s)}_{l\pm m|n}, {\cal F}^{(cc),(sc)}_{l\pm m,k|n}$ used
here contain the coefficients $c_l c_m$, $c_l c_m c_k$, rather
than $c_{l\pm m}, c_{l\pm m} c_k$ as implied by Eq. (\ref{ff}).

Fot the case of SM fermions, the terms with $l = 0, m=0$ in the
Eqs. (\ref{tfd_4d3}) simplify to
\bea{tfd_4d00}
T^{ F SM n }_{\mu \nu} & = & \bar{\psi}^{0}   \left\{
{i\over 4} \biggl[ \gamma_{(\mu} \lrpd_{\nu)} -
2 \eta_{\mu\nu}\gamma^\sigma\lrpd_\sigma \biggr] \mathcal{F}^{(cc)}_{00|n}
 + g \sum_{k}
 \biggl[ \eta_{\mu\nu} \gamma^\rho \hat{B}_\rho^k
  - {1 \over 2} \gamma_{(\mu} \hat{B}_{\nu)}^k \biggr]
  {\cal F}^{(ccc)}_{00k|n}\right\}  \psi^{0}  \nonumber \\
T^{ F SM n }_{\mu 5} & = &  0 \nonumber \\
T^{ F SM n }_{5 5} & = &  \bar{\psi}^{0}   \left\{
{i\over 2} \gamma^\sigma\lrpd_\sigma \mathcal{F}^{(cc)}_{00|n}
 - g \sum_{k}   \gamma^\rho \hat{B}_\rho^k  {\cal F}^{(ccc)}_{00k|n}
 \right\}  \psi^{0}
\eea
where $\psi^0$ is either the left-handed or right-handed part of the fermion,
and $\hat{B} = T^a B^a$ are the corresponding gauge fields.
As expected, in this case the results are equal to the
expressions obtained for the case when matter is restricted
to four dimensions, multiplied by an appropriate form factor. The
Feynman rules for the interactions  of the SM fermions with gravity
can then be obtained directly from \cite{HLZ}.

\subsection{Feynman rules}

\begin{figure}[t!]
\centerline{
   \includegraphics[height=2.3in]{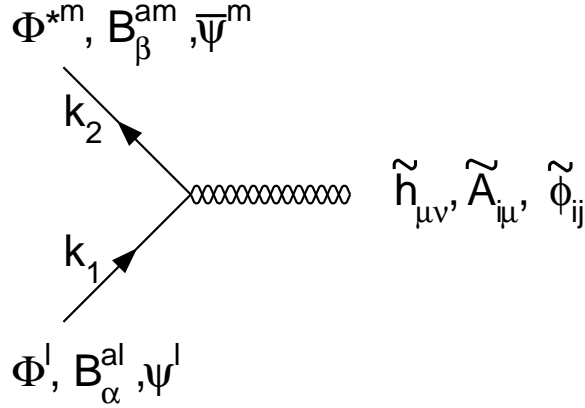}
   }
\caption{ The general interaction vertex between matter
and gravity. The Feynman rules for different types of matter are given in
Eqs. (\ref{vertex}) and Table 1. }
\label{gen_ver}
\end{figure}

It is convenient to express the vertex Feynman rules for the
gravity-matter interactions in terms
of the  components of the energy-momentum tensor in the momentum representation.
For example, define
\bea{mrep}
T^{S \ lmn}_{MN} (k_1,k_2) & = &  <\Phi^m(k_2)| T^{S n}_{MN} | \Phi^l(k_1)>
        \nonumber \\
\eps_1^\al \eps_2^\be T^{V \ lmn}_{MN,\al \be} (k_1,k_2) & = &
<V^m(k_2,\eps_2)| T^{V n}_{MN}
| V^l(k_1,\eps_1)> \nonumber \\
\bar{u}(k_2) T^{F \ lmn}_{MN}(k_1,k_2) u(k_1) & = & < \psi^m(k_2)|
T^{F n}_{MN} | \psi^l(k_1)> \eea Note that only the terms bilinear
in matter fields contribute to these quantities. They  will
describe the radiation of a KK graviton from a scalar, vector
boson or fermion line (with the potential change of the KK
excitation number of the matter as well). From Eq. (\ref{l_4d2})
we then obtain the following vertex interaction rules:
\bea{vertex} \tih^\vn_{\mu \nu} P^m (k_2) P^l (k_1) & :  & -{i
\kappa \over 2}\
T^{P\ lmn}_{\mu \nu} (k_1,k_2) \nonumber \\
\tA^\vn_{i \mu} P^m (k_2) P^l (k_1) & : & i \kappa \
\delta_{i5}\  T^{P \ lmn}_{\mu 5} (k_1,k_2) \nonumber \\
\tp^\vn_{i j} P^m (k_2) P^l (k_1) & : & -{i \kappa \over 2}
\left[ \delta_{ij} \omega
\left( \eta_{\mu \nu} - \frac{p_\mu p_\nu}{m_\vn^2} \right)
   T^{P\ lmn}_{\mu \nu} (k_1,k_2) \right. \nonumber \\
& & \left.  + (\sqrt{2} \delta_{i5} \delta_{j5} - \xi \delta_{ij} )
T^{P\ lmn}_{55}(k_1,k_2)
\right]
\eea
 The convention here is that a KK level-$l$ excitation $P$ with initial momentum
$k_1$ radiates a graviton with momentum $p = k_1-k_2$, becoming
a KK level-$m$ excitation with momentum $k_2$ (as represented pictorially
in Fig. \ref{gen_ver}).

\begin{table}
\begin{center}
\rotatebox{90}{
\begin{tabular}{|c|c|c|c|}
\hline
\hline
& & & \\
 & $T_{\mu \nu}$ & $T_{5 \mu}$ & $T_{55} $\\
& & & \\
\hline
\hline
& & & \\
S
&
${ \displaystyle
(C_{\mu \nu , \rho \si} k_1^\rho k_2^\si + m_\Phi^2 \eta_{\mu \nu})
{\cal F}^{(cc)}_{lm|n} + \frac{m l}{R^2} \eta_{\mu \nu}
{\cal F}^{(ss)}_{lm|n} } $
&
$ i\biggl( k_{2 \mu} {l \over R} {\cal F}^{(sc)}_{lm|n} -
k_{1 \mu} {m \over R} {\cal F}^{(cs)}_{lm|n} \biggr) $
&
$(k_1 k_2 - m_\Phi^2) {\cal F}^{(cc)}_{lm|n} + \frac{m l}{R^2}
{\cal F}^{(ss)}_{lm|n} $\\
& & & \\
\hline
& & & \\
V &
$ \biggl( (m_B^2 - k_1 k_2) C_{\mu \nu , \al \be} -
D_{\mu \nu , \al \be}(k_1,k_2) \biggr) {\cal F}^{(cc)}_{lm|n} $
&
$ i \left[ -{m \over R} D'_{\rho \mu, \al \be} k_1^\rho {\cal F}^{(cs)}_{lm|n}
\right. \ \ \ \ \ \ \ \  $
&
$  -\biggl( D'_{\rho \si, \al \be} k_1^\rho k_2^\si - m_B^2 \eta_{\al \be}
 \biggr) {\cal F}^{(cc)}_{lm|n} $
\\
& $\ \ \ \ + {l m \over R^2} C_{\mu \nu , \al \be} {\cal F}^{(ss)}_{lm|n}$
& $ \ \ \ \ \left. + {l \over R} D'_{\mu \rho, \al \be} k_2^\rho
{\cal F}^{(sc)}_{lm|n} \right] $
&
$  - { ml \over R^2} \eta_{\al \be} {\cal F}^{(ss)}_{lm|n} $ \\
& & & \\
\hline
& & & \\
F &
$ {1 \over 8} C'_{\mu \nu , \rho \si} \ga^\rho (k_1 + k_2)^\si
{\cal G}^{(c) \mp}_{lm|n}  + \eta_{\mu \nu} {\cal A}^\pm_{lmn}$
&
${i \over 8} \biggl( (k_1 + k_2)_\mu {\cal G}^{(s) \pm}_{l,-m|n}
+ \ga_\mu {\cal B}^\pm_{lmn} \biggr) $
&
$ {1 \over 4} (\not{k_1} + \not{k_2}) {\cal G}^{(c) \mp}_{lm|n}
\mp { m_Q \over 2} {\cal G}^{(s) \pm}_{lm|n}
$
\\
& & & \\
\hline
\hline
\end{tabular}
 }
\end{center}
\label{table3} \caption{The bilinear terms of the matter
energy-momentum tensor in momentum representation.}
\end{table}

 The expressions for the quantities defined in Eqs. (\ref{mrep})
are given in Table 1, with the following notations being used (see
also \cite{HLZ}): \bea{Cdef} C_{\mu \nu , \rho \si} & = &
-\eta_{\mu \nu} \eta_{\rho \si} +
\eta_{\mu \rho} \eta_{\nu \si} + \eta_{\mu \si} \eta_{\nu \rho} \nonumber \\
C'_{\mu \nu , \rho \si} & = &  -2 \eta_{\mu \nu} \eta_{\rho \si} +
\eta_{\mu \rho} \eta_{\nu \si} + \eta_{\mu \si} \eta_{\nu \rho} \nonumber \\
D_{\mu\nu,\rho\sigma} (k_1, k_2) & =&  \eta_{\mu\nu}(k_{1\rho} k_{1\sigma}
+k_{2\rho}k_{2\sigma}+k_{1\rho}k_{2\sigma})
-\biggl[\eta_{\nu\sigma}k_{1\mu}k_{1\rho}
+\eta_{\nu\rho}k_{2\mu}k_{2\sigma}
+(\mu\leftrightarrow\nu)\biggr] \nonumber \\
D'_{\rho \si, \al \be} & = & \eta_{ \rho \si} \eta_{\al \be} -
\eta_{\rho \be} \eta_{\al \si} \nonumber \\
{\cal A}^\pm_{lmn} & = & { 1 \over 2 R}
    \left(  l  {\cal G}^{(c)\pm}_{lm|n}  +
     m {\cal G}^{(c)\mp}_{lm|n} \right)
     \pm m_Q  {\cal G}^{(s)\pm}_{lm|n}
    \nonumber \\
{\cal B}^\pm_{lmn} & = & {1 \over R} \left( l  {\cal G}^{(s)\mp}_{l,-m|n}  +
     m  {\cal G}^{(s)\pm}_{l,-m|n}
    \right)
\eea
The upper (lower) sign in the definition of the ${\cal A}, {\cal B}$
factors correspond to the case of doublet (singlet) fermions participating
in the interaction. The Feynman rules for vertices with more that three
particles are given in the Appendix.

The interaction rules for the 0-level KK excitations (the SM
fermions) are easily obtained from Eqs. (\ref{vertex}) by setting
$T_{\mu 5}, T_{55}$ equal to zero. One then obtains the vertex
rules given in \cite{HLZ,GRW}, with one difference: \cite{HLZ,GRW}
make use of the conservation of energy-momentum tensor $p^{\mu}
T_{\mu\nu} = 0$. Therefore they do not have the term proportional
to $p^{\mu} p^{\nu}/m_\vn^2$ in the interaction vertices with the
scalar gravitons. However, energy--momentum
conservation holds only for on-shell particles;
our results are then applicable also
for the case when the SM fermions in the vertex are off-shell.

\section{Phenomenological Implications}

The results obtained in the previous section  are applicable in
all cases when matter and gravity propagate in 5D. For example,
they are also valid in models in which only the gauge bosons or
the Higgs propagate in extra dimensions. These results can also be
used even if we are not in a fat brane scenario, but the matter
fields propagate all the way in the fifth dimension; in that case,
many of the form factors will be zero (due to KK number
conservation), but the general formulas still work.

However, the most  interesting scenario in terms of its
phenomenological consequences may be the UED-type model with
matter on a fat brane. In such a model, in the absence of KK
number violating interactions, the first level KK excitations are
stable \cite{cmn}; or, if the masses of these excitations are
split by loop corrections, the lightest KK particle is stable
\cite{cms}. However, in the presence of gravity, these particles
can decay by radiating KK gravitons \cite{cmn}. The collider
signals are strongly dependent on the decay branching ratios
(gravitational versus electroweak or strong decay) \cite{cmn2}, so
it is necessary to evaluate the decay width to gravitons. This can
be done using the results of the previous section. Moreover, the
interactions of the SM matter with gravity are affected compared
to the case when the matter is restricted to the 4D brane; and
finally, KK number violating graviton exchange can mediate the
production of single KK excitations at hadron or linear colliders.

\subsection{Gravity-mediated decays of KK excitations}

In a fat brane scenario, the gravity-matter interactions do not
respect the KK number conservation rules which hold for matter interactions
in UED
models. Gravity interactions will therefore mediate the decay of
the first level KK excitations of matter (or the lightest KK
particle), which otherwise would be stable.

The decay widths to a single graviton are given below. For the
decay of a KK fermion:
\bea{fer_wid}
\Gamma (q^l \rightarrow q h^\vn) & = & | {\cal F}^c_{l|n} |^2 \
 \frac{\kappa^2}{2\times 384 \pi} \frac{M^3}{x^4}
\left[ \left( 1 - x^2 \right)^4 \left( 2 + 3 x^2 \right)
\right ] \nonumber \\
\Gamma (q^l \rightarrow q A^\vn) & = & | {\cal F}^c_{l|n} |^2 \
 \frac{\kappa^2}{2\times 256 \pi} M^3
\left[ \left( 1 - x^2 \right)^2 \left( 2 +  x^2 \right)
\right ] \times P_{55} \nonumber \\
 \Gamma (q^l \rightarrow q \phi^\vn) & = & | {\cal F}^c_{l|n} |^2 \
 \frac{\kappa^2}{2\times 256 \pi} M^3 \left( 1 - x^2 \right)^2
\left[ c_{11} \frac{( 1 - x^2 )^2}{x^4} \right.
\nonumber \\
& & \left. + 2 c_{12} \frac{ 1 - x^2 }{x^2} + c_{22} \right]
\eea
Here $M$ is the mass of the KK particle $M=l/R$, $m_\vn$ is
the mass of the graviton, and $x = m_\vn/M$. The coefficients
$P_{55}$ and $c_{ij}$ appear because not all $\tA_i, \tp_{ij}$ fields
are independent. To eliminate the spurious
 degrees of freedom we can introduce
the (extra dimensional) polarization vector $e^k_i$ and tensor
$e^s_{ij}$ as in \cite{HLZ}, and we have:
\bea{pol_sums}
P_{55} & = & \sum_{k=1}^{N-1} e_i^k e_j^{k*} \ \delta_{i5} \delta_{j5}
 = 1-\frac{n_5^2}{\vn^2} \nonumber \\
c_{11} & = &  \sum_{s=1}^{N(N-1)/2} e_{ij}^s e_{kl}^{s*} \
\omega^2 \delta_{ij} \delta_{kl} = \omega^2 (N-1) \nonumber \\
c_{12} & = &  \sum_{s=1}^{N(N-1)/2} e_{ij}^s e_{kl}^{s*} \
\omega \delta_{ij}
(\sqrt{2} \delta_{k5} \delta_{l5} - \xi \delta_{kl} ) =
-\frac{2}{N+2} P_{55} \nonumber \\
c_{22} & = &  \sum_{s=1}^{N(N-1)/2} e_{ij}^s e_{kl}^{s*} \
(\sqrt{2} \delta_{i5} \delta_{j5} - \xi \delta_{ij} )
(\sqrt{2} \delta_{k5} \delta_{l5} - \xi \delta_{kl} ) =
\frac{2(N+1)}{N+2} P_{55}^2
\eea

\begin{figure}[t!] 
\centerline{
   \includegraphics[height=3.in]{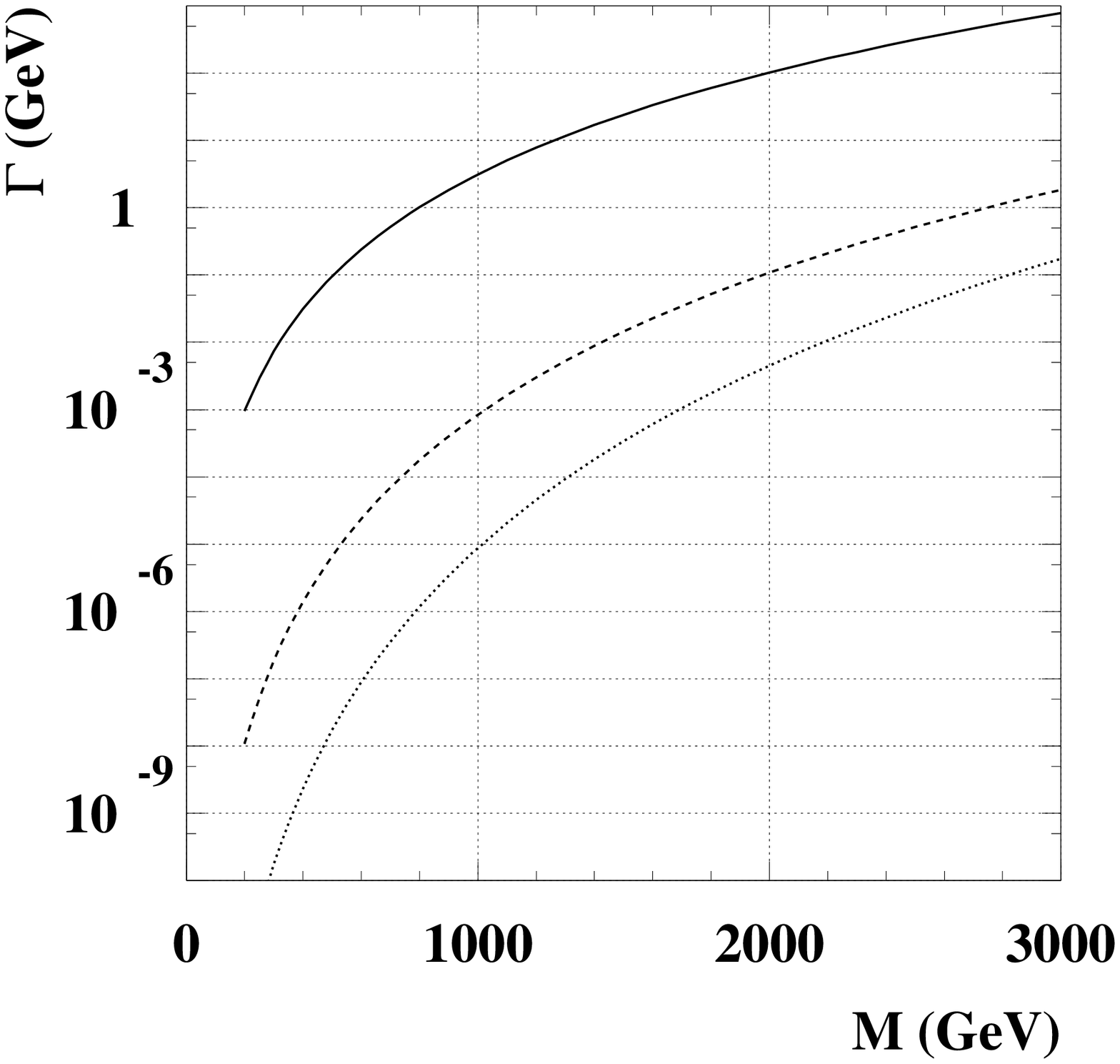}
   \includegraphics[height=3.in]{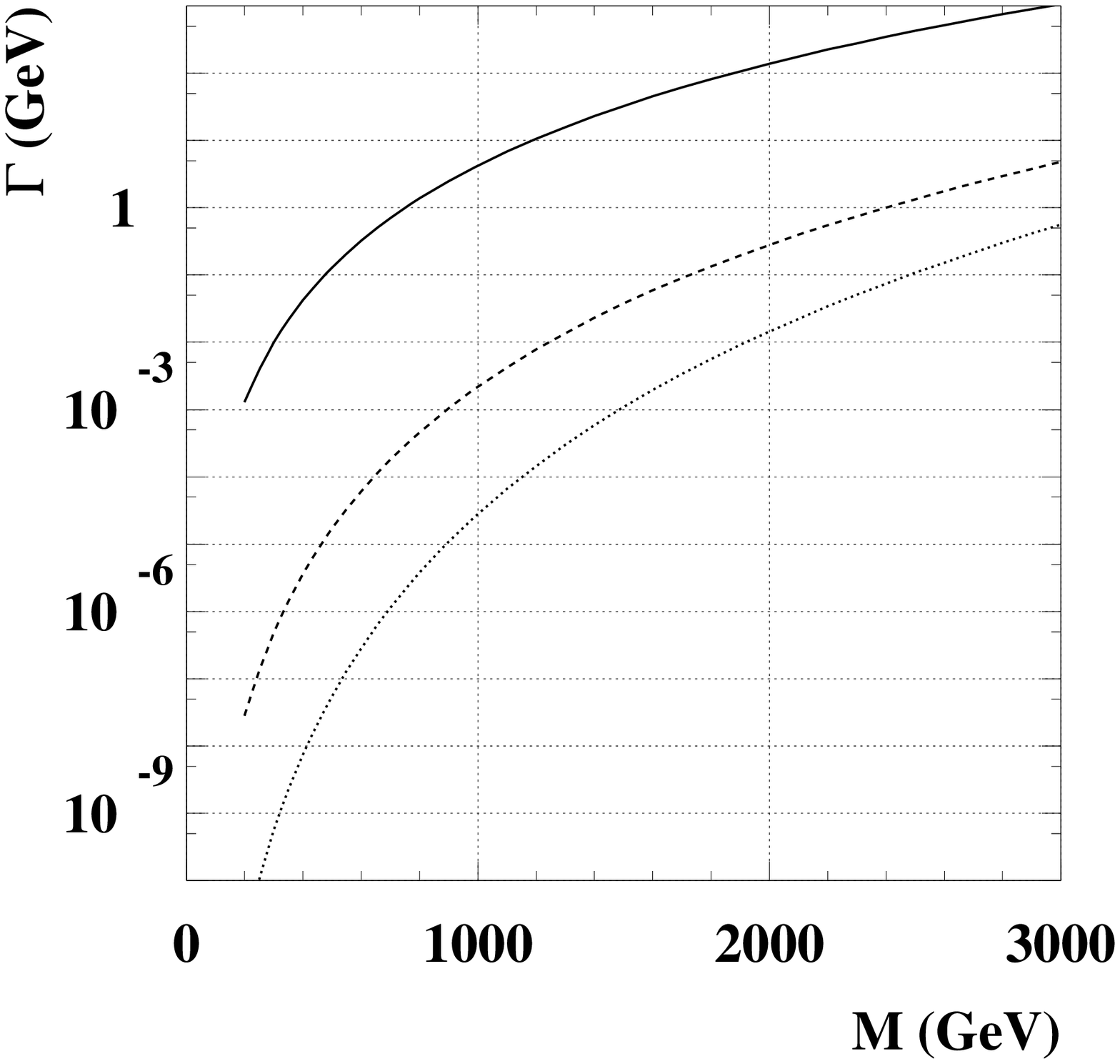}
   }
\caption{ Decay widths for KK fermions (left) and bosons (right)
as a function of the particle mass. Straight lines corresponds to $N=2$
extra dimensions, dashed lines to $N=4$, and dotted lines to $N=6$. }
\label{kkwid}
\end{figure}

For the decay of a KK gauge boson excitation, the following
results are obtained: 
\bea{bos_wid} \Gamma (B^l \rightarrow B
h^\vn) & = & | {\cal F}^c_{l|n} |^2 \
 \frac{\kappa^2}{3\times 96 \pi} \frac{M^3}{x^4}
\left[ \left( 1 - x^2 \right)^3 \left( 1 + 3 x^2 + 6 x^4 \right)
\right ] \nonumber \\
\Gamma (B^l \rightarrow B A^\vn) & = & | {\cal F}^c_{l|n} |^2 \
 \frac{\kappa^2}{3\times 32 \pi} {M^3 \over x^2}
\left[ \left( 1 - x^2 \right)^3 \left( 1 +  x^2 \right)
\right ] \times P_{55} \nonumber \\
 \Gamma (B^l \rightarrow B \phi^\vn) & = & | {\cal F}^c_{l|n} |^2 \
 \frac{\kappa^2}{3\times 32 \pi} M^3 \left( 1 - x^2 \right)^3
\left[ c_{11} \frac{1}{x^4} + 2 c_{12} \frac{ 1 }{x^2} + c_{22}
\right] \eea 
The above formulas apply for the gluon as well as the
electroweak gauge boson excitations, but in the later case the
mass $M$ of the boson acquires a correction coming from the SM
gauge boson mass term.

\begin{figure}[b!] 
\centerline{
   \includegraphics[height=3.in]{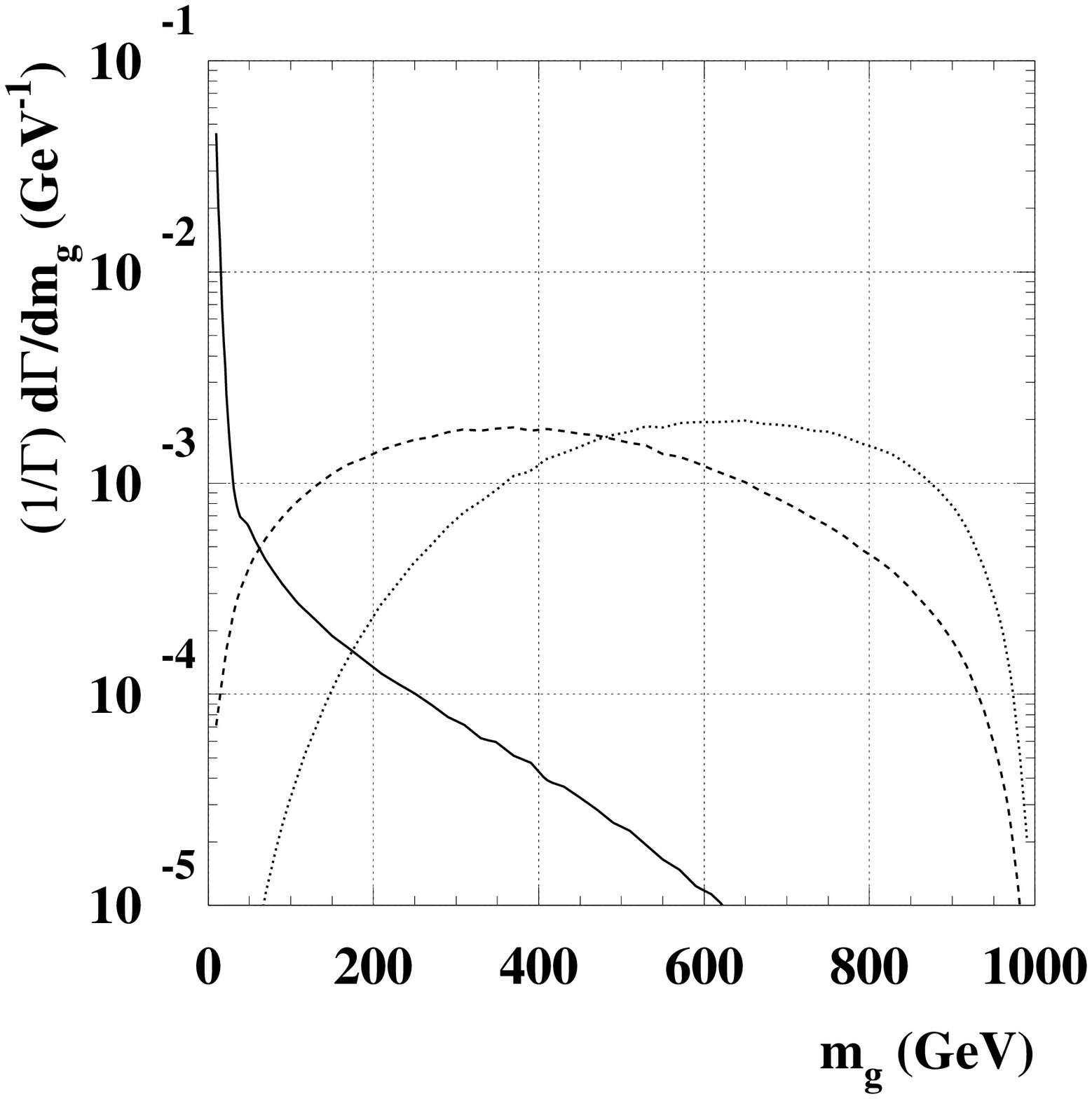}
   \includegraphics[height=3.in]{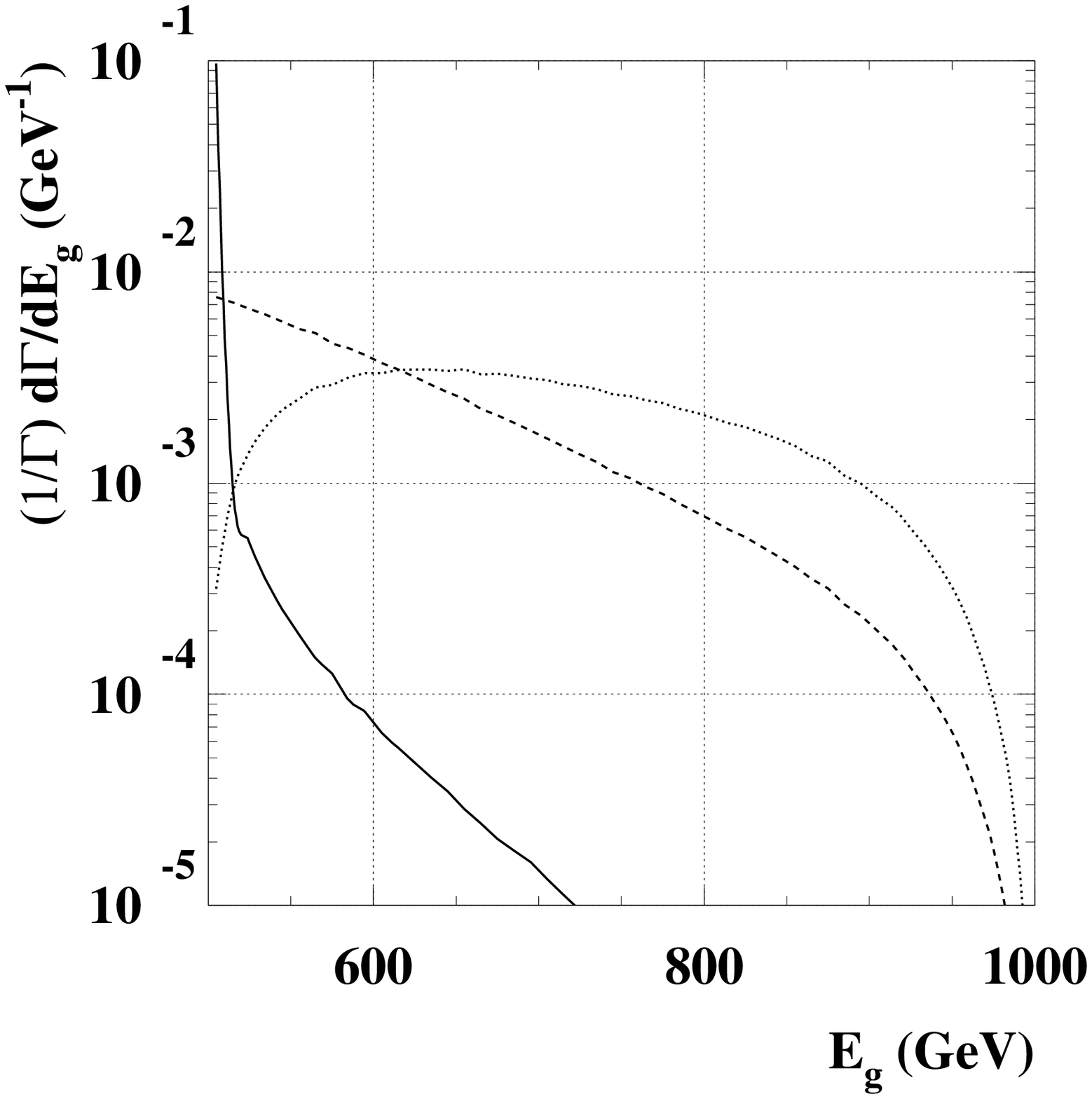}
   }
\caption{ Mass distribution (left) and energy distribution (right)
for the graviton radiated in the decay of one matter KK excitation
with mass 1 TeV. Straight lines corresponds to $N=2$ extra
dimensions, dashed lines to $N=4$, and dotted lines to $N=6$. }
\label{decdist}
\end{figure}

In order to obtain the total gravitational decay width for a
KK excitation, a sum over KK gravitons with mass smaller
than the mass of the decaying particle has to be performed. Since
the masses of the graviton KK excitations are closely spaced,
this sum can be replaced by an integral over the graviton
density of states (for more details, see \cite{HLZ, cmn}).
The results obtained for some
values of the model parameters are presented in Fig. \ref{kkwid}.

The distributions of the mass and energies of gravitons
contributing to the decay of a KK excitations follow the pattern
described in \cite{cmn}. That is, for $N = 2$, the decay is
mediated mostly by light gravitons, and the missing energy (the
energy taken away by the graviton) is about half the particle
mass. For higher $N$, more massive gravitons contribute to the
total decay width, and the missing energy increases to values
close to the KK particle mass, as in Fig. \ref{decdist}.

\subsection{Effects of the fat-brane form factor}

If gravity propagates in large extra dimensions,
SM physics may be affected, either by graviton radiation from the final
(or initial) state of SM processes, or by contributions to these processes
due to virtual exchange of graviton states.
 The fact that the graviton coupling constant
to matter
is proportional to $1/M_{Pl}^2$ is compensated by the large number of gravitons
which can play a role in these interactions. Consequently, there is a
large body of literature dealing with the phenomenological consequences
of KK gravity in large extra dimensions \cite{hewett, rizzo, direct_prod}.

It is natural then to ask how the fat-brane scenario discussed
here affects this phenomenology. The only difference between
the fat-brane and thin-brane cases (where matter is confined in
3 spatial dimensions) is the appearance of the form-factor:
$$ {\cal F}_{0|n_5} = { 1\over \pi R}
\int_0^{\pi R} dy \ \exp \left( 2 \pi i n_5 y \over r\right) $$
which multiplies the relevant interaction vertex.
Since the absolute value of this form factor is smaller than 1,
we can conclude that in this scenario, the width of the brane has an effect of
softening the gravity-SM matter interaction.

In order to better quantify this effect, let us consider the
contribution to 
four-fermion
SM processes due to the exchange of virtual gravitons \cite{HLZ,hewett,rizzo}.
The amplitude for such a contribution can be written as
(see, for example \cite{HLZ}):
$$ {\cal M}(f_1 f_1 \rightarrow G_\vn \rightarrow f_2 f_2) \ = \
{\kappa^2 \over 16 \pi} {i \over s-m_\vn^2 + i\epsilon} \ {\cal A}$$
where ${\cal A}$ is an amplitude which does not depend on the
mass of the graviton being exchanged.
(For the moment we assume that matter is restricted to 4D).
Summing over all graviton states:
$$ {\cal M}(f_1 f_1 \rightarrow \sum_\vn G_\vn \rightarrow f_2 f_2) \ = \
{\kappa^2 \over 16 \pi}\ D(s) \ {\cal A},$$ with \beq{d_def} D(s)
= \sum_\vn {i \over s-m_\vn^2 + i\epsilon}.\eeq The function $D$
can be evaluated by changing the sum into an integral over the
graviton density of states. However, for the number of extra
dimensions in which gravity propagates $N \ge 2$, this integral is
divergent; hence the need to impose an ultraviolet cutoff $M_S$ on
the graviton mass $m_\vn$.

Assuming now that the CM energy of the experiment is much lower than
the cutoff scale $M_S$, we have:
\beq{d_res1} D(s) = \left( r \over 2 \pi \right)^N  V_N \
{- i M_S^{N-2} \over N-2}  \
\left( 1 + {\cal O}\left(s \over M_S^2 \right) \right)
\eeq
where $V_N = 2 \pi^{N/2} / \Gamma (N/2)$ is the volume of the unit sphere
in $N$ dimensions. The above formula is valid for $N>2$;
if $N=2$, the $M_S^{N-2}/(N-2)$ is replaced by $\log(M_S^2/s)$.
Note that there is  a strong dependence on the cutoff mass,
which may mean that the details of physics at the scale where
the theory becomes nonpertubative
will have a large impact on the physics at low energies.\footnote{
For an example of how string-scale physics can affect low scale phenomenology,
see, for example, \cite{peskin}.}

  In the case when matter propagates in one extra dimension,
expression (\ref{d_def}) becomes:
\beq{d_def1}
D_0(s) = \sum_\vn {\cal F}_{0|n_5} \ {i \over s-m_\vn^2 + i\epsilon} \
({\cal F}_{0|n_5})^* 
\eeq
with:
$${\cal F}_{0|n_5} = {i \over \pi}{M \over m_5} \left(
e^{i \pi {m_5 \over M} } - 1 \right)$$
where $M = 1/R$ is the mass of the matter first level KK excitations,
and $m_5 = 2 \pi n_5 /r$ is the part of the graviton mass due to excitation
along the $y$ dimension. Noting that the form factor depends only on $n_5$,
we can integrate along the other extra dimensions to obtain:
\beq{d0_res} D_0(s) = \left( r \over 2 \pi \right)^N V_{N-1} \
\int_0^{M_s} dm_5 |{\cal F}_{0|n_5}|^2
{- i (M_S^2-m_5^2) ^{N-3 \over 2} \over N-3}  \
\left( 1 + {\cal O}\left(s-m_5^2 \over M_S^2 -m_5^2 \right) \right)
\eeq
(here we take $N > 3$).
Due to the $1/m_5$ behavior of the form factor, the main contribution
to the above integral comes from small values of $m_5$; therefore,
the terms of order $m_5^2/M_S^2$ can be neglected. With the change of
variables: $x=\pi m_5/M$, we can rewrite:
\beq{d0_res1}
D_0(s) = \left( r \over 2 \pi \right)^N
V_{N-1} \
{  -i M_S^{N-3} \over N-3}  {M \over \pi} \
\int_0^{\pi M_s/M}  {|e^{i x}-1|^2 \over x^2} dx .
\eeq
The integral in the above expression has an asymptotic limit
$\pi$ (when $M_S \gg M$).
Therefore, the effective strength of the four-fermion
interaction operator is reduced in the fat-brane scenario by a factor
$ \sim M/ M_S$.

From a phenomenological point of view, this will make the observation
of virtual graviton exchange effects at colliders more difficult.
Therefore, the limits set in \cite{hewett,rizzo} on the string scale $M_D$ can
be somewhat lowered. We estimate that this softening effect will
be less visible in processes with radiation of real gravitons, since
in that case the energy of the collider provides a natural cutoff scale for
the sum over graviton excitations which is less than $M$.\footnote{
Conversely, if the collider energy is larger than $M$, KK excitations
of matter can be observed directly.}

From a theoretical point of view, this behavior is interesting
because it reduces the dependence of the strength of the effective
interactions mediated by gravity at low energies on the details of
the theory at the scale $M_D$. This can be seen from the fact that
in this model, where matter propagates in one extra dimension, for
$N=2$ the cutoff scale can be let go to infinity (the sum over KK graviton
propagators is
convergent), while for $N=3$ the dependence on the cutoff scale is
logarithmic. In a model where the brane is `fat' in all $N$ extra
dimensions, sums like (\ref{d_def1}) are finite for any value of
$N$. In this manner, the non-zero thickness of the brane provides
a softening factor for the coupling of gravity to matter (for a
different mechanism, see \cite{soft_brane}), in effect replacing
the string scale $M_D$ by the inverse thickness of the brane $M$
as a cutoff factor.

\subsection{Production of single KK excitations}

Finally, we will consider the effect which gravity interactions
may have in the production of a single matter KK excitation.
Due to KK number conservation in UED models, matter KK excitations
can be produced only in pairs. However, the gravity-matter interaction
does not obey the KK number conservation rule, therefore graviton
exchange may mediate the production of a single KK excitation at
linear or hadron colliders.
 The relevant propagator sum for such a process is:
$$ D_1(s) = \sum_\vn {\cal F}_{0|n_5} \ {i \over -m_\vn^2 } \
({\cal F}^c_{1|n_5})^*.$$
Since we assume that the experiment energy is much smaller than the string
scale $M_D$ (or the cutoff scale $M_S$), we have neglected it in the above sum,
which therefore describes both $s$-channel and $t$-channel contributions.
 Integrating over the
graviton density of states like in the previous section we then obtain:
$$
D_1(s) = \left( r \over 2 \pi \right)^N
V_{N-1} \
{ -i M_S^{N-3} \over N-3}  {M \over \pi}
{- 2 i \sqrt{2} \over \pi^2} \
\int_0^{\pi M_s/M}  {\sin x \over 1-x^2/\pi^2} dx .
$$
The integral above has the asymptotic value $\pi$SinIntegral($\pi$)
$\simeq$ 5.81 .

The amplitude for the production of a single KK excitation is therefore
of order $\kappa^2 s D_1 \sim s^2 M/M_D^5$
(we take the cutoff scale to be $M_S \simeq M_D$). On the other hand,
the amplitude for the production of two KK states is of order $g^2$,
where $g$ is the coupling constant for the force which mediates
the interaction (strong at a hadron collide, electroweak at a linear collider).
Then it can be expected that the ratio of cross-sections for single KK
production versus double KK production should be of order
$ (s^2 M/g^2 M_D^5 )^2$ which seems to be quite small. However,
the fact that greater CM energy is needed to produce two KK excitations has
also to be taken into account. If $2M > \sqrt{s}$ at a linear collider,
production of double KK excitations cannot take place. Also, at a hadron
collider, the effective luminosity decreases rather fast with $s$.


\begin{figure}[t!] 
\centerline{
   \includegraphics[height=3.in]{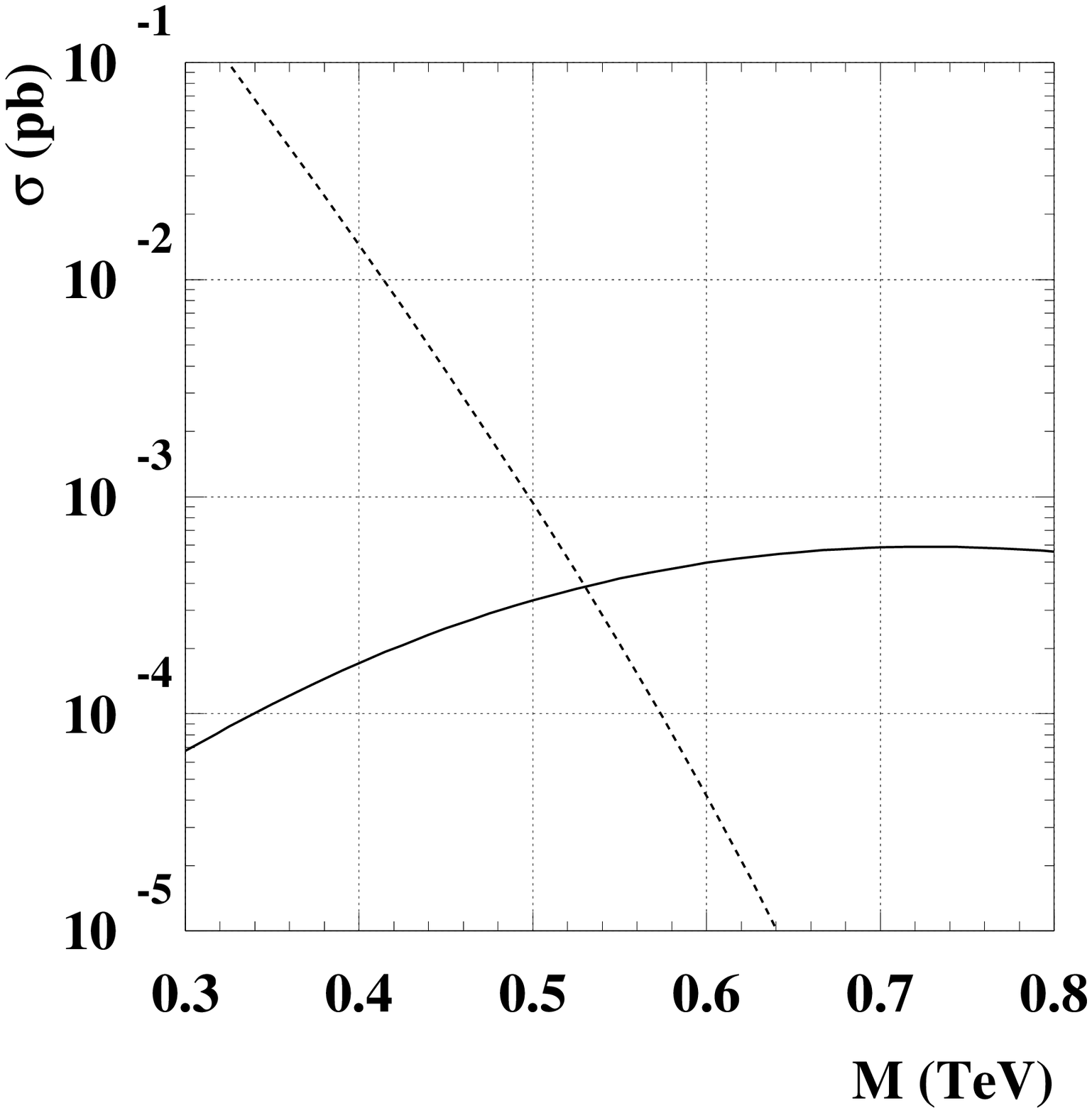}
   \includegraphics[height=3.in]{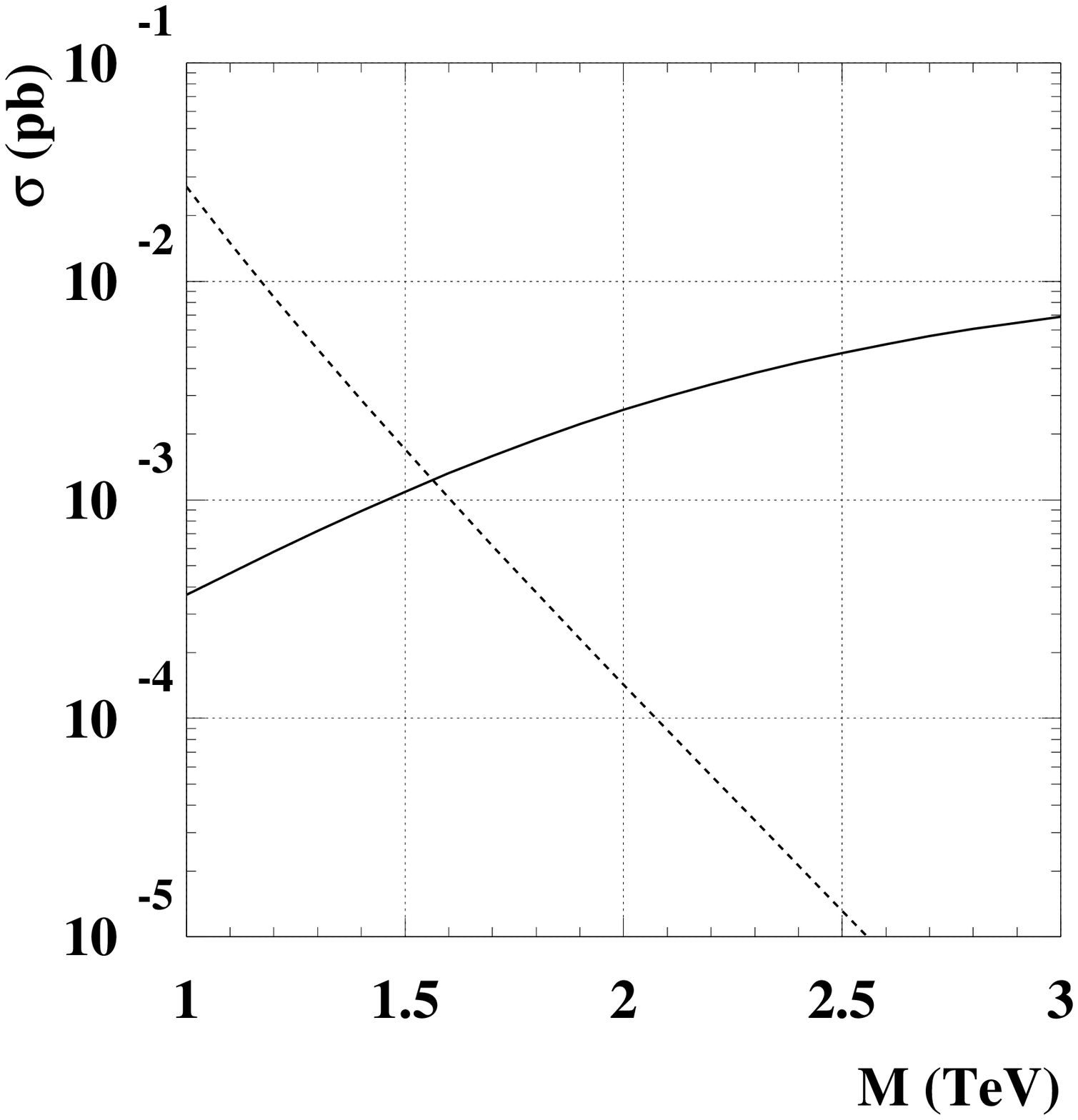}
   }
\caption{ $q \bar{q} \rightarrow b \bar{b}^*$ (straight line)
and  $q \bar{q} \rightarrow b^* \bar{b}^*$ (dotted line) production rates
at Tevatron Run II (left) and LHC (right). }
\label{singleK}
\end{figure}

For illustration, we present in Fig. \ref{singleK} the production
cross-section for a single KK excitation of the $b$ quark
(process mediated by the $s$-channel exchange of a KK tower of gravitons),
and the production cross-section
for the double KK state $b^* \bar{b}^*$. We take $M_D = 2$ TeV for the
Tevatron cross-section, and $M_D = 5$ TeV for the LHC
cross-section\footnote{For $M_D$ , we use the definition:
 $M_{Pl}^2 = (r/2 \pi)^{N+2} M_D^N $.}. Note that,
up to values of $M$ of about 1/3 of the total collider CM energy,
the cross-section for single KK production increases with mass. This is
due to the $s^2 M/M_D^5 \sim (M/M_D)^5$ behavior of the amplitude.
(For higher values of $M$, the decrease in effective luminosity will
bring the cross-section down).
This behavior may allow for the observation of a single KK state even if
the compactification scale $M$ is too large for the production
of two excited states.
 However, the cross-section for single KK production
 will be large enough to be observable
  only in the case when $M_D$ is not much
larger than $M$, since every doubling of the string scale will reduce
this cross-section  by about three orders of magnitude.

\section{Conclusions}

If there are large extra dimensions, gravity may have an important
role to play in the phenomenology of present day colliders. Almost
all analyses of the gravitational effects so far concentrate on
scenarios with matter being restricted to the 4-dimensional brane.
However, gravity can also have an important role to play in models
where matter also propagate into  extra dimensions. In this paper,
we analyze the interactions of matter with gravity in such models.

For the gravitational field, we use the linearized lagrangian
which describes small perturbations around a flat background
metric. The reduction to KK modes and the gauge fixing for the
gravitons is done in a manner identical to \cite{HLZ}. The results
obtained for the gravity-matter interaction are valid for a large
class of models where matter fields propagate into one extra
dimension. For the sake of specificity, we consider a case where
all matter fields propagate a reduced distance along the fifth
dimension (UED on the fat brane scenario). We then compute the
energy-momentum tensor of the matter scalar, vector and fermion
fields in five dimensions,  perform the reduction to KK modes, and
derive the general Feynman rules for interaction between the KK
excitations of matter and gravity.

We then make use of these results for some phenomenological analysis. First,
we compute the gravitational decay widths of first level KK excitations
of matter. If such excitations are produced at hadron collider, their decay
modes will play an important role in their detection and identification, as
discussed in \cite{cmn2}. Second, we discuss the effect which the thickness
of the brane has on the interactions of SM matter with gravity. We find
out that having a fat brane reduces the strength of the effective gravitational
coupling by a factor proportional to the superposition of the wave functions
of matter and gravity along the fifth direction. Perhaps more interesting, we
find out that the thickness of the brane can act as an effective cutoff scale
in the evaluation of sums over propagators of virtual KK gravitons.

 Finally, we consider the case of production of single KK excitations of
quarks and gluons at hadron colliders. Due to KK number conservation
in UED-type models, KK excitations of matter are usually produced in pairs,
thus requiring a large CM energy even for relatively small compactification
scale. However, the gravitational interactions break KK number conservation
(in models
with matter on a fat brane), therefore they can mediate single KK production.
We show that the cross-section for such a process is reasonably large, as long
as the string scale $M_D$ is not much larger than the mass of the matter
KK excitations.

\vspace{1.cm}
{\Large \bf Acknowledgments}
\vspace{0.5cm}

This work was supported in
part by the U.S. Department of Energy Grant Numbers
DE-FG03-98ER41076, DE-FG02-01ER45684, and DE-FG02-91ER40685.

\vspace{1.cm}
{\Large {\bf Appendix}}

\begin{figure}[t!] 
\centerline{
   \includegraphics[width=5.in]{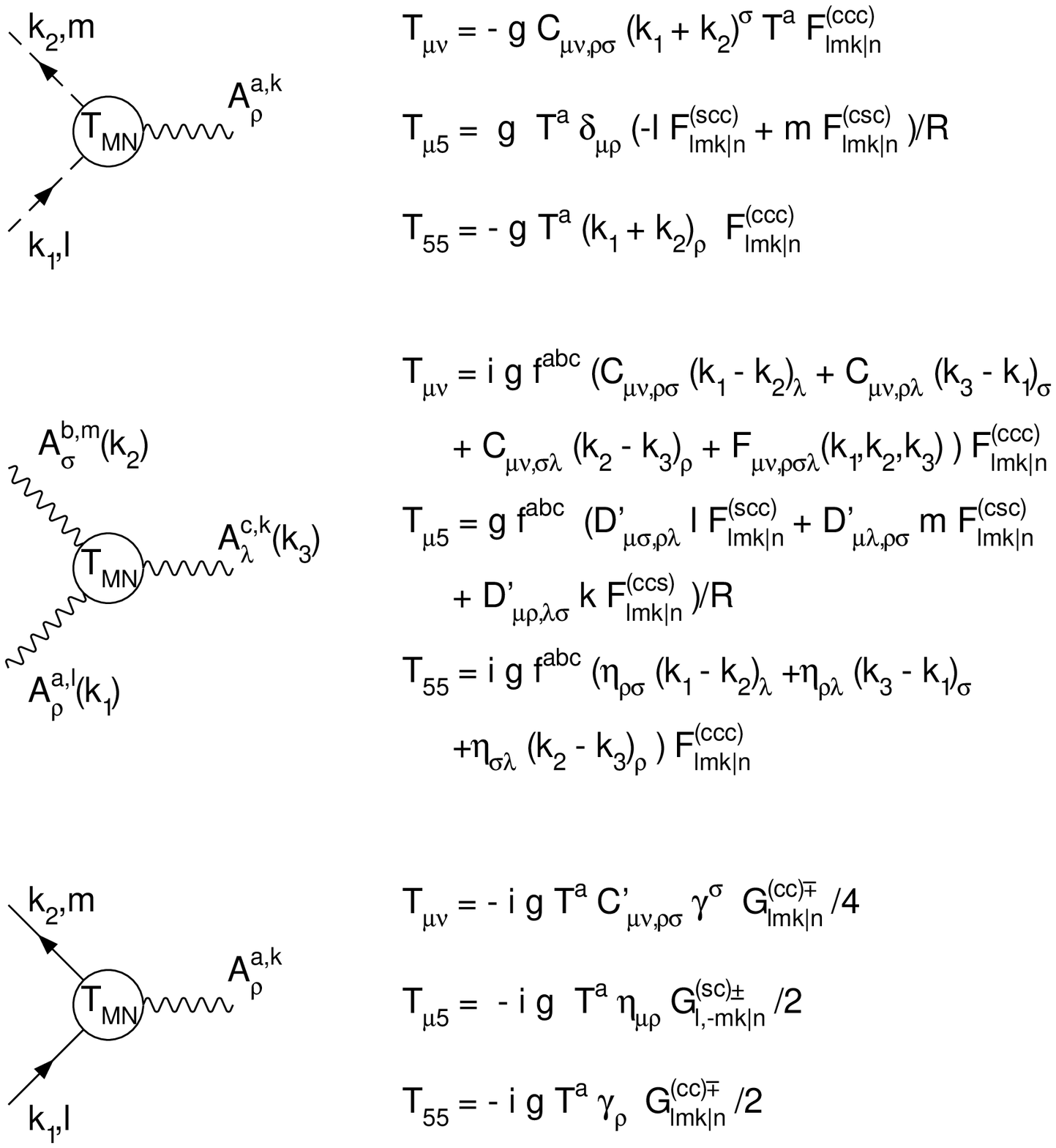}
   }
\caption{ The trilinear components of the matter energy-momentum
tensor in momentum representation. Symbols are defined in the
text.} \label{grav3}
\end{figure}

\vspace{0.5cm}
We present here the components of the matter energy-momentum tensor which
are trilinear and quadrilinear in the matter fields. For simplicity, these
components are given in momentum representation. The Feynman rules for
the gravity interactions can then be easily derived (as in Eqs. (\ref{vertex})).

\begin{figure}[t!] 
\centerline{
   \includegraphics[width=5.in]{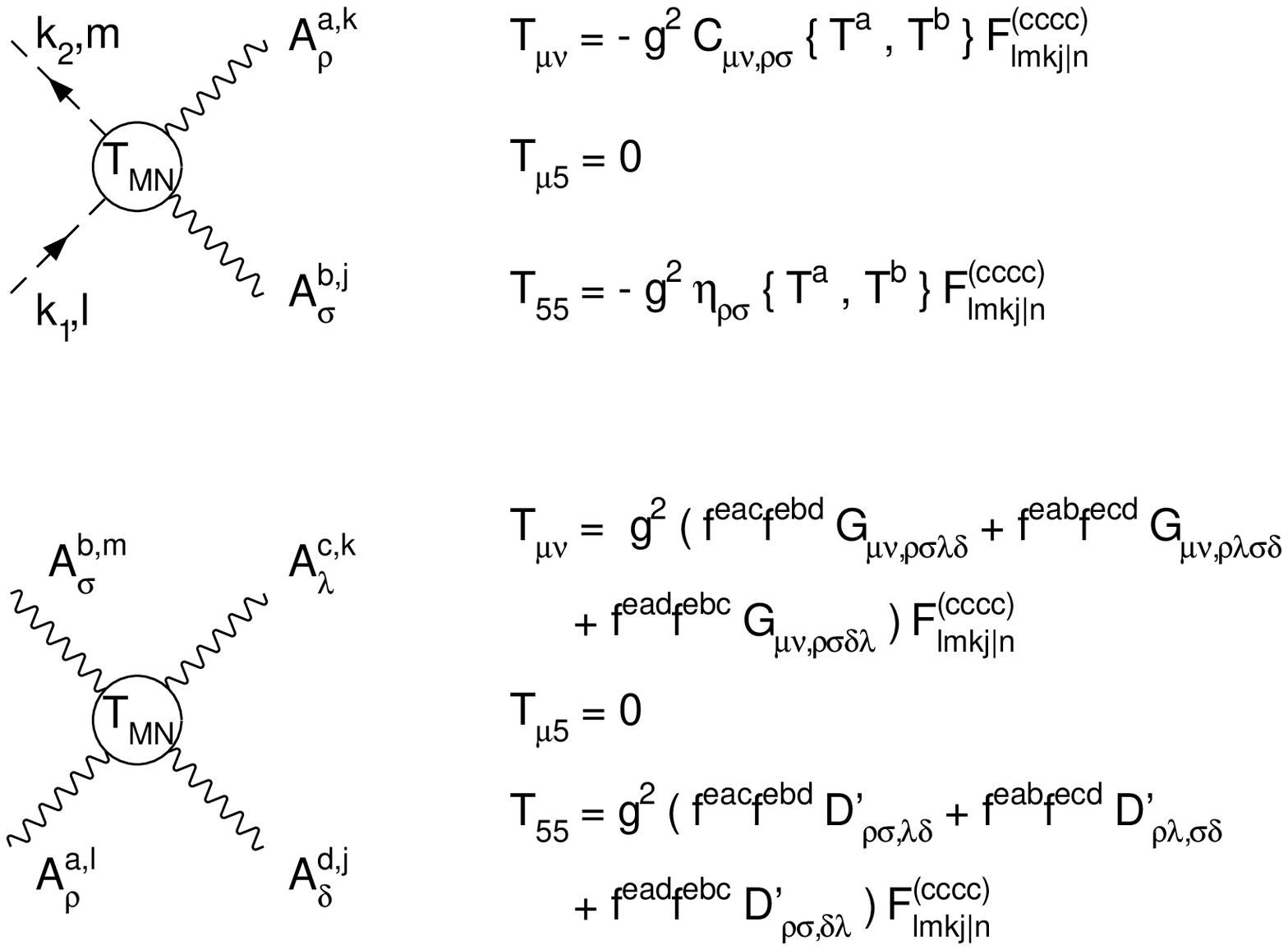}
   }
\caption{ The quadrilinear components of the matter
energy-momentum tensor in momentum representation. Symbols are
defined in the text.} \label{grav4}
\end{figure}

 Fig. \ref{grav3} contains the trilinear terms in the matter energy-momentum
tensor, while Fig. \ref{grav4} contains the quadrilinear terms.
 The following notations are used:
\bea{fg_def}
F_{\mu\nu,\ro\si\la}(k_1,k_2,k_3) & = &
\eta_{\mu\ro}\eta_{\si\la}(k_2-k_3)_{\nu} +
\eta_{\mu\si}\eta_{\ro\la}(k_3-k_1)_{\nu} \nonumber \\
& & + \eta_{\mu\la}\eta_{\ro\si}(k_1-k_2)_{\nu} +
( \mu \leftrightarrow \nu ) \nonumber \\
G_{\mu\nu,\ro\si\la\de} & = & \eta_{\mu\nu} D'_{\ro\si,\la\de} +
\biggl( \eta_{\mu\ro}\eta_{nu\de}\eta_{\la\si}
+ \eta_{\mu\la}\eta_{nu\si}\eta_{\ro\de} \nonumber \\
& & - \eta_{\mu\ro}\eta_{nu\si}\eta_{\la\de}
 - \eta_{\mu\la}\eta_{nu\de}\eta_{\ro\si} + ( \mu \leftrightarrow \nu )
 \biggr).
\eea The quantities $C_{\mu\nu,\ro\si}, C'_{\mu\nu,\ro\si},
D_{\mu\nu,\ro\si}(k_1,k_2)$, and $D'_{\ro\si,\la\de}$ are defined
in Eqs. (\ref{Cdef}), $g$ is the gauge coupling constant,
$f^{abc}$ are the structure constants of the gauge group and $T^a$
are the generators of the representation of the gauge group
associated with the corresponding matter fields. Also, $j,k,l,m$
are the KK indices of the matter fields. In the definition of the
fermion--fermion--gauge-boson vertex, the upper sign holds for the
case of doublet fields $Q^l, Q^m$ while the lower sign holds for
the case of singlet fields $q^l, q^m$.

\end{document}